\documentclass[aps,prd,preprintnumbers,superscriptaddress,nofootinbib,twocolumn]{revtex4}%

\usepackage{graphicx}
\usepackage{color}
\usepackage{amssymb}  
\usepackage{amsmath}
\def\({\biggl(}
\def\){\biggr)}
\def\[{\biggl[}
\def\]{\biggr]}
\def\mpl{M_{\rm Pl}}
\def\beq{\begin{equation}}
\def\eeq{\end{equation}}

\allowdisplaybreaks
 \def\be   {\begin{equation}}   \def\ee   {\end{equation}}
 \def\ba  {\begin{eqnarray}}   \def\ea  {\end{eqnarray}}
\begin{document}

\title{The fate of non-diagonalizable interactions in quasidilaton theory}

\author{Rampei~Kimura}
\affiliation{Center for Cosmology and Particle Physics, Department of Physics, New York University, New York, NY, 10003
}

\begin{abstract}
It has been shown that the spherically symmetric solutions in a subclass of quasidilaton theory 
are stable against all degrees of freedom and does not even exhibit superluminal propagation. 
These solutions can be found by switching off scalar-tensor interactions, which can not be removed by a local transformation.
In this paper, we extend the analysis to quasidilaton theory, including non-diagonalizable scalar-tensor interactions.
We show that all solutions inside the Vainshtein radius are problematic : 
the scalar mode in massive graviton suffers from gradient instabilities, the vector mode are infinitely strongly coupled vector perturbations, or the Vainshtein mechanism is absent.
\end{abstract}
%

\maketitle

\section{Introduction}
Modifying gravitational theories typically introduces an additional degrees of freedom, which mediates a fifth force, and it is tightly constrained by solar system experiments \cite{Will:2005aa}. 
Therefore, modified gravity requires a mechanism to restore general relativity at short distances to pass these tests (for a review see \cite{Joyce:2014aa}).
One of the reliable mechanism, which reduces to general relativity at short distances with an excellent accuracy, could be the Vainshtein mechanism \cite{Vainshtein:1972aa}, which is originally found in the context of the van Dam-Veltman-Zhakarov discontinuity \cite{vDV:1970aa,Z:1970aa} in Fiertz-Pauli theory \cite{FP:1939aa} and 
can be also found in the decoupling limit of Dvali-Gabadadze-Polatti model \cite{Dvali:2000hr,Luty:2003vm} and Galileon theory \cite{Nicolis:2008in}.

Such a mechanism can be seen in ghost-free massive gravity called de Rham-Gabadadze-Tolley (dRGT) massive gravity \cite{deRham:2010ik,Rham:2011aa}, whose mass term consists of an infinite number of potentials to eliminate the Boulware-Deser (BD) ghost \cite{Boulware:1972aa} (for the proof of the absence of BD ghost in full theory, see \cite{Deffayet:2013aa,Hassan:2011hr,Kugo:2014aa,Mirbabayi:2011aa}). In the decoupling limit, which enable us to capture physics within a certain scale,
the dynamics of five polarization modes can be described within a certain class of scalar(-vector)-tensor theories. 
In the subparameter space such that the scalar-tensor interactions can be diagonalized by a local transformation (this type of theory is called "restriced galileon" in the literature), 
spherically symmetric assumption allows the stable asymptotically cosmological solutions against perturbations for all modes although an asymptotically Minkowski solution is forbidden due to the appearance of helicity-0 ghost of massive graviton \cite{Berezhiani:2013aa}. 
However the instabilities arise when the non-diagonalizable scalar-tensor interactions are included in the theory \cite{Berezhiani:2013ab}. 

This fact would motivate us to look for stable solutions in another type of massive gravity.
The minimal extension of introducing the scalar degree of freedom in massive gravity is so-called quasidilaton theory \cite{DAmico:2012aa} (for a further extension, see \cite{Felice:2013aa}). 
This quasidilaton $\sigma$ can be introduced by a global symmetry, $\phi^a \to e^\alpha \phi^a$ and $\sigma \to \sigma  - \alpha\mpl $, where $\alpha$ is a constant and $\phi^a$ is the St\"{u}ckelberg field.
We can now ask the same question as in dRGT theory :  Is there any stable spherically symmetric solutions in quasidilaton theory? 
Some of the answer has been already solved in \cite{Gabadadze:2014ab}, which investigated spherically symmetric solutions in the decoupling limit of quasidilaton theory within a certain parameter space where non-diagonalizable  interactions is absent. The authors found that the asymptotically cosmological solutions are free of ghost instabilities, gradient instabilities, and superluminalities although the asymptotically Minkowski solution can not be allowed as with the case of restricted galileon.
Therefore the aim of this paper is to  extend the analysis in \cite{Gabadadze:2014ab} by taking into account the entire parameter space of quasidilaton theory.

This paper is organized as follows. 
In Sec. II, we describe the decoupling limit theory of quasidilaton and derive all relevant equations.
In Sec III, we find spherically symmetric solutions and its consequences to perturbations in the simplest case : no shift-symmetric Horndeski terms for $\sigma$.
In Sec IV, we extend Sec III by adding shift-symmetric Horndeski terms.
Sec V is devoted to the summary. 

We adopt the signature $(-,+,+,+)$ for the metric throughout this work, 
and use the following shortcut notation,  
$\varepsilon^{\mu\alpha\rho\sigma}\varepsilon^{\nu\beta}_{~~\rho\sigma}\Pi_{\mu\nu}\Pi_{\alpha\beta}
\equiv\varepsilon\varepsilon\Pi\Pi$, 
$\varepsilon_\mu ^{~\gamma\alpha\rho}\varepsilon_{\nu \gamma}^{~~\beta\sigma}
\Pi_{\alpha\beta}\Pi_{\rho\sigma}\equiv\varepsilon_\mu\varepsilon_\nu\Pi\Pi$, 
$(B^2)^\mu_\nu\equiv B^\mu_{~\alpha} B^\alpha_{~\nu}$,
$\varepsilon\varepsilon B\partial A \equiv \varepsilon_{\mu_1 \mu_2 \mu_3 \mu_4}\varepsilon^{\nu_1 \nu_2 \mu_3 \mu_4} B^{\mu_1}_{~\nu_1} \partial_{\nu_2} A^{\mu_2}$, 
and so on.

\section{The theory}
\label{sec:2}
In this paper we consider the decoupling limit of quasidilaton theory.
The decoupling limit is defined as $\mpl \to 0$, $m \to 0$, $\Lambda =(\mpl m^2)^{1/3}={\rm fixed}$, and 
$T_{\mu\nu}/\mpl={\rm fixed}$. 
All six degrees of freedom (five polarization modes in massive graviton and quasidilaton) can be decomposed into the massless tensor $h_{\mu\nu}\equiv g_{\mu\nu}-\eta_{\mu\nu}$, 
the massless vector $A_\mu$,  and the two scalars $\pi$ and $\sigma$, 
via the relation $\phi^a = \delta^a_\mu x^\mu - \eta^{a\mu}A_\mu/(\mpl m)-\eta^{a\mu} \partial_\mu \pi/(\mpl m^2)$.
The scalar-tensor Lagrangian in the decoupling limit, up to the relevant energy scale $\Lambda$, can be written as \cite{Gabadadze:2014ab}
\begin{eqnarray}
	&&{\cal L}
	=-{1\over 4} h^{\mu\nu}{\cal E}_{\mu\nu}^{\alpha\beta}h_{\alpha\beta}
	-h^{\mu\nu} \[{1\over 4}\varepsilon_\mu\varepsilon_\nu \Pi 
	-{\alpha \over 4 \Lambda^3}\varepsilon_\mu\varepsilon_\nu \Pi\Pi\nonumber\\
	&&~~~~~
	-{\beta\over 2\Lambda^6}\varepsilon_\mu\varepsilon_\nu \Pi\Pi\Pi
	-{\xi_2 \over 2 \Lambda^3}\varepsilon_\mu\varepsilon_\nu \Sigma\Sigma
		-{\xi_4 \over 2\Lambda^6}\varepsilon_\mu\varepsilon_\nu \Sigma\Sigma\Sigma
	\]\nonumber\\
	&&~~~~~
	+\sigma\biggl[4\alpha_5 \Lambda^3+ \gamma_0\varepsilon\varepsilon\Pi
	+{\gamma_1\over \Lambda^3}\varepsilon\varepsilon\Pi\Pi
	+{\gamma_2\over \Lambda^6}\varepsilon\varepsilon\Pi\Pi\Pi\nonumber\\
	&&~~~~~
	+{\gamma_3\over \Lambda^9}\varepsilon\varepsilon\Pi\Pi\Pi\Pi
	-{\omega \over 12}\varepsilon \varepsilon \Sigma
	-{\xi_1 \over 6\Lambda^3}\varepsilon \varepsilon \Sigma\Sigma\nonumber\\
	&&~~~~~
	-{\xi_3 \over 4\Lambda^6}\varepsilon \varepsilon \Sigma\Sigma\Sigma
	-{\xi_5 \over 10\Lambda^9}\varepsilon \varepsilon \Sigma\Sigma\Sigma\Sigma
	\biggr]\nonumber\\
	&&~~~~~
	+{1 \over 2M_{\rm Pl}}h^{\mu\nu}T_{\mu\nu},
\label{LagrangianDL}
\end{eqnarray}
where ${\cal E}_{\mu\nu}^{\alpha\beta}$ is the Einstein operator, 
$\Pi_{\mu\nu} = \partial_\mu \partial_\nu \pi$, $\Sigma_{\mu\nu} = \partial_\mu \partial_\nu \sigma$, $T_{\mu\nu}$ is the energy momentum tensor, and 
$\alpha$, $\alpha_5$, $\beta$, $\omega$, $\gamma_i$, and $\xi_i$ are parameters\footnote{
Note that $\gamma_i$ are related to the parameters through
$\gamma_0=(3 -4\alpha_5)/6$, 
$\gamma_1=-(2+\alpha+2\alpha_5)/2$,
$\gamma_2=2(1+\alpha-\alpha_5)/3$, and 
$\gamma_3=-(1+\alpha+\alpha_5)/6$.
}.
The Lagrangian ${\cal L}$ is invariant under diffeomorphism transformation $h_{\mu\nu} \to h_{\mu\nu} + 2\partial_{(\mu}\zeta_{\nu)}$ and internal Galilean transformations for $\pi$ and $\sigma$, 
$\partial_\mu \pi \to \partial_\mu \pi + c_\mu$ and $\partial_\sigma \pi \to \partial_\sigma \pi + d_\mu$.
The terms involving the parameter $\xi_i$ are obtained from shift-symmetric Horndeski Lagrangian \cite{Horndeski:1974aa,Deffayet:2009aa,Deffayet:2013ab} for $\sigma$ field\footnote{The derivation of the decoupling limit of Horndeski theory was investigated in \cite{Koyama:2013aa}. }, and setting all $\xi_i=0$ yields the original quasidilaton theory proposed in \cite{DAmico:2012aa}.
If $\beta=\xi_4=0$, which is the case investigated in \cite{Gabadadze:2014ab}, 
the action can be recasted into a subclass of bi-galileon action \cite{Hinterbichler:2010aa,Padilla:2010aa,Padilla:2010ab} via a local transformation, 
$h_{\mu\nu} \to h_{\mu\nu} +\pi \eta_{\mu\nu} -(\alpha/\Lambda^3) \pi \Pi_{\mu\nu} -(2\xi_2/\Lambda^3)\sigma \Sigma_{\mu\nu} $. 

The vector Lagrangian is independent of the tensor modes $h_{\mu\nu}$ and quasidilaton $\sigma$, 
but couples with the scalar $\pi$ \cite{Gabadadze:2013aa,Ondo:2013wka},  
\begin{eqnarray}
	&&{\cal L}_{A}=-{1\over 4} 
	\biggl[
	\Lambda^3 \varepsilon\varepsilon BB + 2(1-\alpha)\varepsilon\varepsilon BB \Pi
	\nonumber\\
	&&~~~~~~~~~~~~~~~
		-{\alpha+6\beta \over \Lambda^3}\varepsilon\varepsilon BB\Pi\Pi
	+\varepsilon\varepsilon B^2 \Pi
	\nonumber\\
	&&~~~~~~~~~~~~~~~
	-{ \alpha \over \Lambda^3} \varepsilon\varepsilon B^2 \Pi\Pi
	-{2 \beta \over \Lambda^3}\varepsilon\varepsilon B^2 \Pi\Pi\Pi
	\nonumber\\
	&&~~~~~~~~~~~~~~~
	+2\Lambda^{3/2} \varepsilon\varepsilon B \partial A 
		-{4 \alpha \over \Lambda^{3/2}}\varepsilon\varepsilon B \partial A \Pi 
	\nonumber\\
	&&~~~~~~~~~~~~~~~
	-{12 \beta \over \Lambda^{9/2}}\varepsilon\varepsilon B \partial A \Pi \Pi
	\biggr],
	\label{LagrangianSV}
\end{eqnarray}
where $B_{\mu\nu}$ is an auxiliary non-dynamical anti-symmetric tensor, 
and the Lagrangian ${\cal L}_{A}$ is invariant under $U(1)$ gauge transformation, $A_\mu \to A_\mu + \partial_\mu \chi$.

In this paper we consider the general ansatz for spherically symmetric background, 
\begin{eqnarray}
 h_{00} = h(r), \qquad h_{ij}= f(r) \delta_{ij},
\end{eqnarray}
for tensor modes, and 
\begin{eqnarray}
\pi(t,x) &=& {a \over 2}\Lambda^3 t^2 + \pi(r), \nonumber\\
\sigma(t,x) &=& {b \over 2} \Lambda^3 t^2 + \sigma(r),
\end{eqnarray}
for scalar modes. 
Here we assume that the vector mode can be ignored at the background level, $A_{\mu} = 0$.
Then the equation of motion for $h_{\mu\nu}$ yields two independent equations for $f$ and $a$, 
\begin{eqnarray}
	r f' &=& - \frac{2M}{\mpl r} + \Lambda^3 r^2 \[\lambda-\alpha\lambda^2-2\beta\lambda^3
	\nonumber\\
	&&~~~~~~~~~~~~~~~~~~~~~~
	-2\xi_2\lambda_\sigma^2-2\xi_4 \lambda_\sigma^3\],\nonumber\\
	r h' &=& - \frac{2M}{\mpl r}
		+ \Lambda^3 r^2  \[a-(1+2a \alpha)\lambda-6a \beta\lambda^2-2\beta\lambda^3
	\nonumber\\
	&&~~~~~~~~~~~~~~~~~
	-4b \xi_2 \lambda_\sigma-6b\xi_4 \lambda_\sigma^2-2\xi_4 \lambda_\sigma^3\],
	\label{gEOM}
\end{eqnarray}
where the prime denotes the derivative with respect to $r$, and we defined the dimensionless variables,
\begin{eqnarray}
\lambda\equiv {\pi' \over \Lambda^3 r,} \qquad \lambda_\sigma\equiv {\sigma' \over \Lambda^3 r},
\end{eqnarray}
 and the Vainshtein scale, 
\begin{eqnarray}
r_*\equiv \({M  \over 4\pi M_{\rm Pl}^2 m^2}\)^{1/3}.
\end{eqnarray}
The $\pi$ and $\sigma$ equations of motion can be compactly given by
\begin{eqnarray}
	&& \sum_{n,m=0}^{n+m \leq 5} A^{(i)}_{n,m}\lambda^n \lambda_\sigma^m=B^{(i)}\left({r_* \over r} \right)^3,
	\label{EOM}
\end{eqnarray}
where $i$ denotes $\pi$ and $\sigma$, corresponding to $\pi$-equation and $\sigma$-equation respectively, 
and the coefficients $A^{(i)}_{n,m}$ are independent of $\lambda$ and $\lambda_\sigma$, listed in the appendix A. 
On the other hand the coefficients  $B^{(i)}$ are functions of $\lambda$ and $\lambda_\sigma$, which are given by
\begin{eqnarray}
		B^{(\pi)}&=&2(1+2a\alpha+12a\beta\lambda+6\beta\lambda^2),\nonumber\\
		B^{(\sigma)} &=& 18 \xi_4 \lambda_\sigma (2b + \lambda_\sigma ).
\end{eqnarray}
The crucial differences from the case $\beta=0$ are $\lambda$ and $\lambda_\sigma$ dependences on $B^{(i)}$. These terms could potentially yield the new type of solutions inside the Vainshtein radius, and 
we will see this fact in the next section.
On the other hand, the asymptotically Minkowski solution, which can be obtained by equating the relevant terms at the linear regime,  is still given by the one in \cite{Gabadadze:2014ab}, however $\pi$ is unfortunately the ghost mode around this background. 
Therefore the solutions outside the Vainshtein radius should be at least nontrivial cosmological solutions, 
which are $\lambda,\lambda_\sigma = {\rm const}$ although stability conditions need to be investigated.

\section{Vainshtein solution without shift-symmetric Horndeski terms}
Let us consider the simplest case, $\xi_i=0$, corresponding to the absence of shift-symmetric Horndeski terms for $\sigma$ field.  
In this case, $\sigma$ equation is linear in $\lambda_\sigma$, thus it is analytically solvable for $\lambda_\sigma$. 
Then $\lambda^5$ and $\lambda^2 (r_*/r)^3$ in the master equation for $\pi$ are the dominant components well inside the Vainshtein radius, and we have the approximate solutions,
\begin{eqnarray}
	\lambda \simeq x_1 {r_* \over r}, 
	\qquad \lambda_\sigma \simeq y_1 \({r_* \over r}\)^3
\label{sol1}
\end{eqnarray}
where $x_1$ and $y_1$ are constants, 
\begin{eqnarray}
	x_1&=& \pm \biggr|\frac{\beta \omega}{-2(\gamma_2-4a\gamma_3)^2-\beta^2\omega}\biggr|^{1/3},\\
	\qquad y_1&=&\frac{2x_1^3(\gamma_2-4a\gamma_3)}{\omega}.
\end{eqnarray}
Then the equations of motion for $f$ and $h$ can be rewrited as 
\begin{eqnarray}
	f' = -(1+x_1^3 \beta) {2M \over \mpl r^2}+ {\cal O}\({1 \over r}\), \\
	h' = -(1+x_1^3 \beta) {2M \over \mpl r^2}+ {\cal O}\({1 \over r}\).
\end{eqnarray}
One can clearly see that the contribution from helicity-0 mode cannot be screened in this solution, i.e., 
there is no Vainshtein mechanism. 
However we can redefine the Plank mass (or Newton's constant) as the one that we observe at short distances such as ${\bar M}_{\rm Pl} \equiv \mpl/ (1+x_1^3 \beta)$ so that the leading order in the gravitational potential agrees with Newtonian one. However, the parametrized post-Newtonian expansion gives the different result as one in general relativity, which could be tightly constrained by solar system experiments. 
Therefore, we require $x_1^3 \beta \ll 1$.
Now let's take a look at the scalar perturbations around this background. 
The quadratic Lagrangian for the scalar perturbations are given by
\begin{eqnarray}
	{\cal L}^{(2)} &\supset&  {\cal A}(\partial_t \phi)^2 +{\cal B}(\partial_t \psi)^2+{\cal C}(\partial_t \phi)(\partial_t \psi), \nonumber\\
	& =& {\cal A}\(\partial_t \phi + {{\cal C} \over 2 {\cal A} }\partial_t \psi\)^2+
	\({\cal B}- {{\cal C}^2 \over 4 {\cal A} }\)(\partial_t \psi)^2.
	\label{quadraticL1}
\end{eqnarray}
where $\phi(t,x)$ and $\psi(t,x)$ are the perturbations of $\pi$ and $\sigma$, respectively, and the coefficients ${\cal A}$ and ${\cal B}$ are given by
\begin{eqnarray}
{\cal A} \simeq -12 \gamma_3 x_1^2 y_1\({r_* \over r}\)^5, \quad 
{\cal B} \simeq  {\omega \over 2}, 
\end{eqnarray}
and ${\cal C} \propto (r_*/r)^2$, making $ {{\cal C}^2 / 4 {\cal A} } $ the small correction in Eq.~(\ref{quadraticL1}).
Thus we require $-\gamma_3 x_1^2 y_1>0$ to avoid ghost for (diagonalized) $\pi$ field and $\omega>0$ for $\sigma$ field. 
This feature is completely contrast to the case $\beta=0$, which is $ {{\cal C}^2 / 4 {\cal A} } \gg {\cal B}$ inside the Vainshtein radius, leading to the ghost mode for the scalar perturbations.
Next we would like to evaluate the sound speed of $\pi$ perturbations.
Following to the method in \cite{Gabadadze:2014ab},  one can find that the radial and angular sound speeds are related to the kinetic coefficient ${\cal A}$ at the leading order inside the Vainshtein radius, 
\begin{eqnarray}
 c_r^2 &\simeq& 3 c_\Omega ^2\nonumber \\
 &\simeq&  - \frac{1}{32\omega}\(\frac{2(\gamma_2-4 a \gamma_3)^2 + \beta^2 \omega}{\beta \gamma_3}\)^2 \({r \over r_*}\)^6{\cal A}.
\end{eqnarray}
Since $\omega>0$, one cannot eliminate the leading order of $c_r^2$ by
any combination of the parameters.
 Thus we arrive at  gradient instabilities in both radial and angular direction for $\pi$ perturbations.

Another type of solution can be obtained by setting the right-hand side in Eq.~(\ref{EOM}) to be zero, 
$B^{(\pi)}=0$, or explicitly,
\begin{eqnarray}
	1+2a\alpha+12a\beta\lambda+6\beta\lambda^2=0.
	\label{condition1}
\end{eqnarray}
In this case, $\lambda$ is constant everywhere, and the $\pi$ force is successfully screened since the additional contribution from the scalar is $\delta(f'), ~\delta(h')  \simeq {\rm const}$
. 
Then the leading contribution to the gradient energy in $\pi$'s perturbation is given by
\begin{eqnarray}
	{\cal L}^{(2)} \simeq-3\beta (a+\lambda) \({r_* \over r}\)^3 \[ 2 (\partial_r \phi)^2-(\partial_\Omega \phi)^2\].
	\label{quadraticL}
\end{eqnarray}
If $\lambda \neq -a $, one of the squared sound speeds of radial and angular part is always negative, leading to the gradient instability in the $\pi$-sector.
This feature due to non-diagonalizable interactions is already addressed in the context of the decoupling limit of dRGT massive gravity \cite{Berezhiani:2013ab} and Horndeski theory \cite{Koyama:2013aa}. Therefore we require $\lambda = -a $ at short distances. 
However the kinetic term of the vector perturbations are given by
\begin{eqnarray}
	{\cal L}_{A}^{(2)} = \frac{1-2\alpha\lambda-6\beta\lambda^2}{4+2a-2\lambda} (\partial_t {\bf A})^2 + \cdot\cdot\cdot,
\end{eqnarray}
where we set the gauge condition $\nabla \cdot {\bf A}=0$ and $A_\mu=(0, {\bf A})$\footnote{
The auxiliary tensor $B_{\mu\nu}$ is already integrated out in this expression.	
}.
One can read off the numerator of the coefficient is the same combination that we imposed in (\ref{condition1}) if $\lambda = -a $.
Thus the vector perturbations is infinitely strongly coupled in this solution.

It should be noted that the scalar field $\pi$ is the Lorentz invariant form $\pi = (a/2)  x^\mu x_\mu$ if $\lambda = -a $, 
and in this case the stable self-accelerating solution can be found in a broad parameter range \cite{Gabadadze:2014aa}. However, the condition (\ref{condition1}) is not the case of the stable one because the vector perturbations are infinitely strongly coupled. 
If the solution asymptotically approaches this stable de Sitter space-time, the solution inside the Vainshtein radius is described by (\ref{sol1}) and 
the $\pi$ perturbation suffers from the gradient instabilities discussed in the above. 
Therefore the self-accelerating solution found in \cite{Gabadadze:2014aa} is problematic at short distances.

\section{Vainshtein solution with shift-symmetric Horndeski terms}
In the case $\beta=0$, the shift-symmetric Horndeski terms for $\sigma$ field are crucial for the stable perturbations and subluminal propagations \cite{Gabadadze:2014ab}. 
In this section we include the shift-symmetric Horndeski terms in the case $\beta \neq 0$.
In the presence of these terms, i.e., $\xi_i \neq 0$, the equations for $\pi$ and $\sigma$ are coupled quintic equations, thus we assume the following ansatz inside the Vainshtein radius,  
\begin{eqnarray}
	\lambda \simeq x_1 {r_* \over r}, \qquad \lambda_\sigma \simeq y_1  {r_* \over r}.
\end{eqnarray}
Then the equations of motion for $\pi$ and $\sigma$ gives the same equations for $x_1$ and $y_1$,
\begin{eqnarray}
	1+ \beta x_1^3+\xi_4 y_1^3 =0.
	\label{constraint}
\end{eqnarray}
By using these solutions, the metric fluctuations can be written as 
\begin{eqnarray}
	f' &=& -(1+ \beta x_1^3+\xi_4 y_1^3) {2M \over \mpl r^2}+ {\cal O}\({1 \over r}\),\nonumber\\
	h' &=& -(1+ \beta x_1^3+\xi_4 y_1^3) {2M \over \mpl r^2}+ {\cal O}\({1 \over r}\).
\end{eqnarray}
One can clearly see that the leading term are canceled due to Eq.~(\ref{constraint}),  
and both scalar modes completely screen the term from massless graviton.
Then the leading terms in these metric fluctuations are $f' \sim h' \sim 1/r$;
therefore, this solution can not even reproduce the Newtonian profile. 
We disregard this solution for this obvious reason.

As in the previous section, we have the other type of solution, $\lambda,~\lambda_\sigma=$ constant everywhere in space. For $\xi_i \neq 0$ case, we further impose the condition $B^{(\sigma)}=0$ in addition to $B^{(\pi)}=0$, which completely eliminates the source terms in both $\pi$ and $\sigma$ equations.
Then these conditions translate into Eq.~(\ref{condition1}) and 
\begin{eqnarray}
\lambda_\sigma(2b+\lambda_\sigma)=0.
\label{condition2}
\end{eqnarray}
As one can see, introducing the shift-symmetric Horndeski does not change anything about 
the condition (\ref{condition1}) and  the gradient energy of $\pi$'s perturbations (\ref{quadraticL}) as well as the vector perturbations (because the vector mode only couples with $\pi$). Therefore we conclude that the vector perturbations are infinitely strongly coupled even in the presence of Horndeski terms. 
Furthermore, the leading contribution to the gradient energy for $\sigma$'s perturbation is given by
\begin{eqnarray}
{\cal L}^{(2)}_\sigma=-3\xi_4 (b+\lambda_\sigma) \({r_* \over r}\)^3 \[ 2 (\partial_r \psi)^2-(\partial_\Omega \psi)^2\].
\label{quadraticLs}
\end{eqnarray}
We have $\lambda_\sigma=0,-2b$ from Eq.~(\ref{condition2}), which means that
the $\sigma$ perturbations always suffer from gradient instabilities.

\section{Summary}
In this paper we investigated the possibility of stable spherically symmetric solutions in the decoupling limit of quasidilaton theory in the whole parameter space.
We showed that the presence of non-diagonalizable scalar-tensor interactions contains the following serious problems.
One of the solutions inside the Vainshtein radius
can not be allowed due to the appearance of the gradient instabilities for $\pi$ perturbations.
For the solution, which does not depend on the source term in equations of motions, 
the extra degrees of freedom $\pi$ and $\sigma$ can be successfully screened, but the vector perturbations are infinitely strongly coupled for any parameter space. 
We confirmed that these conclusions can not be evaded in the case of inclusion of shift-symmetric Horndeski interactions for $\sigma$ field.
One of the solutions does not even have the Newtonian gravitational potential at the leading order due to the cancellation with the contributions from the scalar modes. The other solution, whose equations of motion is independent of the source term, encounters the same problem as in the case of the absence of Horndeski terms.
Therefore,  the case $\beta=0$ found in \cite{Gabadadze:2014ab} is 
the only consistent quasidilaton theory, which are free of ghosts, tachyons, gradient instability, and superluminality and is not ruled out by solar system experiments.



\acknowledgments 
We would like to thank Gregory Gabadadze for very
useful discussions.
R.K. is supported in part by JSPS Postdoctoral Fellowships for Research Abroad. 

\appendix

\section{Coefficients in equations of motion for $\pi$ and $\sigma$}
In this appendix we summarize the coefficients of equations of motion defined in (\ref{EOM}).
$A^{(\pi)}_{n,m}$ are given by
	\begin{eqnarray}
	A^{(\pi)}_{0,0}&=&-a+ 4 b \gamma_0, \nonumber\\
	A^{(\pi)}_{1,0}&=& 3+6a\alpha+8b\gamma_1,\nonumber\\
	A^{(\pi)}_{2,0}&=& -6(\alpha+a\alpha^2-4a\beta-2b\gamma_2),\nonumber\\
	A^{(\pi)}_{3,0}&=& 2(\alpha^2-4\beta-20a \alpha\beta+8b\gamma_3),\nonumber\\
	A^{(\pi)}_{4,0}&=& -60a \beta^2,\nonumber\\
	A^{(\pi)}_{5,0}&=& -12\beta^2,\nonumber\\
	A^{(\pi)}_{0,1}&=& -4(3\gamma_0-2a\gamma_1+b\xi_2),\nonumber\\
	A^{(\pi)}_{0,2}&=& 2(2\xi_2+2a\alpha\xi_2-3b\xi_4),\nonumber\\
	A^{(\pi)}_{0,3}&=& 2(1+2a\alpha)\xi_4,\nonumber\\
	A^{(\pi)}_{1,1}&=& -8(2\gamma_1-3a\gamma_2-b\alpha\xi_2),\nonumber\\
	A^{(\pi)}_{1,2}&=& -4(\alpha\xi_2-6a\beta\xi_2-3b\alpha\xi_4),\nonumber\\
	A^{(\pi)}_{1,3}&=& -24a\beta\xi_4,\nonumber\\
	A^{(\pi)}_{2,1}&=& -12(\gamma_2-4a\gamma_3-2b\beta\xi_2),\nonumber\\
	A^{(\pi)}_{2,2}&=& -36b\beta\xi_4,\nonumber\\
	A^{(\pi)}_{2,3}&=& -12\beta\xi_4,
	\end{eqnarray}
	and $A^{(\sigma)}_{n,m}$ are given by
\begin{eqnarray}
	A^{(\sigma)}_{0,0}&=& 4\alpha_5 + 6 a \gamma_0 - b \omega,\nonumber\\
	A^{(\sigma)}_{1,0}&=& -6(3\gamma_0-2a\gamma_1- b\xi_2),\nonumber\\
	A^{(\sigma)}_{2,0}&=& -3(4\gamma_1-6a\gamma_2+b \alpha \xi_2),\nonumber\\
	A^{(\sigma)}_{3,0}&=& -6(\gamma_2-4 a \gamma_3+2b\beta \xi_2),\nonumber\\
	A^{(\sigma)}_{0,1}&=& 3(\omega-2b\xi_1 + 2 a \xi_2),\nonumber\\
	A^{(\sigma)}_{0,2}&=& 3(2\xi_1-12b\xi_2^2-6b\xi_3+3a\xi_4),\nonumber\\
	A^{(\sigma)}_{0,3}&=& 6(2\xi_2^2+\xi_3-20b\xi_2\xi_4-2b\xi_5),\nonumber\\
	A^{(\sigma)}_{0,4}&=& -90 b\xi_4^2,\nonumber\\
	A^{(\sigma)}_{0,5}&=& -18\xi_4^2,\nonumber\\
	A^{(\sigma)}_{1,1}&=& -6(2\xi_2+a \alpha \xi_2-3b \xi_4),\nonumber\\
	A^{(\sigma)}_{1,2}&=& -3(3+2 a \alpha)\xi_4,\nonumber\\
	A^{(\sigma)}_{2,1}&=& 3(\alpha \xi_2 -12 a \beta \xi_2 -2b \alpha \xi_4),\nonumber\\
	A^{(\sigma)}_{2,2}&=& -54 a \beta \xi_4,\nonumber\\
	A^{(\sigma)}_{3,1}&=& -36b\beta \xi_4,\nonumber\\
	A^{(\sigma)}_{3,2}&=& -18\beta \xi_4,
\end{eqnarray}
and any other coefficients $A^{(\pi, \sigma)}_{n,m}$ are zero.

\bibliography{references}

\begin{thebibliography}{31}
\expandafter\ifx\csname natexlab\endcsname\relax\def\natexlab#1{#1}\fi
\expandafter\ifx\csname bibnamefont\endcsname\relax
  \def\bibnamefont#1{#1}\fi
\expandafter\ifx\csname bibfnamefont\endcsname\relax
  \def\bibfnamefont#1{#1}\fi
\expandafter\ifx\csname citenamefont\endcsname\relax
  \def\citenamefont#1{#1}\fi
\expandafter\ifx\csname url\endcsname\relax
  \def\url#1{\texttt{#1}}\fi
\expandafter\ifx\csname urlprefix\endcsname\relax\def\urlprefix{URL }\fi
\providecommand{\bibinfo}[2]{#2}
\providecommand{\eprint}[2][]{\url{#2}}

\bibitem[{\citenamefont{Will}(2005)}]{Will:2005aa}
\bibinfo{author}{\bibfnamefont{C.~M.} \bibnamefont{Will}},
  \bibinfo{journal}{Living Rev.Rel.} \textbf{\bibinfo{volume}{9}},
  \bibinfo{pages}{3} (\bibinfo{year}{2005}).

\bibitem[{\citenamefont{Joyce et~al.}(2014)\citenamefont{Joyce, Jain, Khoury,
  and Trodden}}]{Joyce:2014aa}
\bibinfo{author}{\bibfnamefont{A.}~\bibnamefont{Joyce}},
  \bibinfo{author}{\bibfnamefont{B.}~\bibnamefont{Jain}},
  \bibinfo{author}{\bibfnamefont{J.}~\bibnamefont{Khoury}}, \bibnamefont{and}
  \bibinfo{author}{\bibfnamefont{M.}~\bibnamefont{Trodden}}
  (\bibinfo{year}{2014}), \eprint{arXiv:1407.0059}.

\bibitem[{\citenamefont{Vainshtein}(1972)}]{Vainshtein:1972aa}
\bibinfo{author}{\bibfnamefont{A.~I.} \bibnamefont{Vainshtein}},
  \bibinfo{journal}{Phys. Lett.} \textbf{\bibinfo{volume}{B39}},
  \bibinfo{pages}{393} (\bibinfo{year}{1972}).

\bibitem[{\citenamefont{H.~van Dam}(1970)}]{vDV:1970aa}
\bibinfo{author}{\bibfnamefont{M.~G.~V.} \bibnamefont{H.~van Dam}},
  \bibinfo{journal}{Nucl. Phys.} \textbf{\bibinfo{volume}{B22}},
  \bibinfo{pages}{397} (\bibinfo{year}{1970}).

\bibitem[{\citenamefont{Zakharov}(1970)}]{Z:1970aa}
\bibinfo{author}{\bibfnamefont{V.~I.} \bibnamefont{Zakharov}},
  \bibinfo{journal}{JETP.Lett.} \textbf{\bibinfo{volume}{12}},
  \bibinfo{pages}{312} (\bibinfo{year}{1970}).

\bibitem[{\citenamefont{M.~Fierz}(1939)}]{FP:1939aa}
\bibinfo{author}{\bibfnamefont{W.~P.} \bibnamefont{M.~Fierz}},
  \bibinfo{journal}{Proc. R. Soc.} \textbf{\bibinfo{volume}{A173}},
  \bibinfo{pages}{211} (\bibinfo{year}{1939}).

\bibitem[{\citenamefont{Dvali et~al.}(2000)\citenamefont{Dvali, Gabadadze, and
  Porrati}}]{Dvali:2000hr}
\bibinfo{author}{\bibfnamefont{G.}~\bibnamefont{Dvali}},
  \bibinfo{author}{\bibfnamefont{G.}~\bibnamefont{Gabadadze}},
  \bibnamefont{and} \bibinfo{author}{\bibfnamefont{M.}~\bibnamefont{Porrati}},
  \bibinfo{journal}{Phys.Lett.} \textbf{\bibinfo{volume}{B485}},
  \bibinfo{pages}{208} (\bibinfo{year}{2000}).

\bibitem[{\citenamefont{Luty et~al.}(2003)\citenamefont{Luty, Porrati, and
  Rattazzi}}]{Luty:2003vm}
\bibinfo{author}{\bibfnamefont{M.~A.} \bibnamefont{Luty}},
  \bibinfo{author}{\bibfnamefont{M.}~\bibnamefont{Porrati}}, \bibnamefont{and}
  \bibinfo{author}{\bibfnamefont{R.}~\bibnamefont{Rattazzi}},
  \bibinfo{journal}{JHEP} \textbf{\bibinfo{volume}{0309}}, \bibinfo{pages}{029}
  (\bibinfo{year}{2003}).

\bibitem[{\citenamefont{Nicolis et~al.}(2009)\citenamefont{Nicolis, Rattazzi,
  and Trincherini}}]{Nicolis:2008in}
\bibinfo{author}{\bibfnamefont{A.}~\bibnamefont{Nicolis}},
  \bibinfo{author}{\bibfnamefont{R.}~\bibnamefont{Rattazzi}}, \bibnamefont{and}
  \bibinfo{author}{\bibfnamefont{E.}~\bibnamefont{Trincherini}},
  \bibinfo{journal}{Phys.Rev.} \textbf{\bibinfo{volume}{D79}},
  \bibinfo{pages}{064036} (\bibinfo{year}{2009}).

\bibitem[{\citenamefont{de~Rham and Gabadadze}(2010)}]{deRham:2010ik}
\bibinfo{author}{\bibfnamefont{C.}~\bibnamefont{de~Rham}} \bibnamefont{and}
  \bibinfo{author}{\bibfnamefont{G.}~\bibnamefont{Gabadadze}},
  \bibinfo{journal}{Phys.Rev.} \textbf{\bibinfo{volume}{D82}},
  \bibinfo{pages}{044020} (\bibinfo{year}{2010}).

\bibitem[{\citenamefont{de~Rham et~al.}(2011)\citenamefont{de~Rham, Gabadadze,
  and Tolley}}]{Rham:2011aa}
\bibinfo{author}{\bibfnamefont{C.}~\bibnamefont{de~Rham}},
  \bibinfo{author}{\bibfnamefont{G.}~\bibnamefont{Gabadadze}},
  \bibnamefont{and} \bibinfo{author}{\bibfnamefont{A.~J.}
  \bibnamefont{Tolley}}, \bibinfo{journal}{Phys.Rev.Lett.}
  \textbf{\bibinfo{volume}{106}}, \bibinfo{pages}{231101}
  (\bibinfo{year}{2011}).

\bibitem[{\citenamefont{Boulware and Deser}(1972)}]{Boulware:1972aa}
\bibinfo{author}{\bibfnamefont{D.~G.} \bibnamefont{Boulware}} \bibnamefont{and}
  \bibinfo{author}{\bibfnamefont{S.}~\bibnamefont{Deser}},
  \bibinfo{journal}{Phys. Rev.} \textbf{\bibinfo{volume}{D6}}
  (\bibinfo{year}{1972}).

\bibitem[{\citenamefont{Deffayet et~al.}(2013)\citenamefont{Deffayet, Mourad,
  and Zahariade}}]{Deffayet:2013aa}
\bibinfo{author}{\bibfnamefont{C.}~\bibnamefont{Deffayet}},
  \bibinfo{author}{\bibfnamefont{J.}~\bibnamefont{Mourad}}, \bibnamefont{and}
  \bibinfo{author}{\bibfnamefont{G.}~\bibnamefont{Zahariade}},
  \bibinfo{journal}{JCAP} \textbf{\bibinfo{volume}{1301}}, \bibinfo{pages}{032}
  (\bibinfo{year}{2013}).

\bibitem[{\citenamefont{Hassan and Rosen}(2012)}]{Hassan:2011hr}
\bibinfo{author}{\bibfnamefont{S.~F.} \bibnamefont{Hassan}} \bibnamefont{and}
  \bibinfo{author}{\bibfnamefont{R.~A.} \bibnamefont{Rosen}},
  \bibinfo{journal}{Phys.Rev.Lett.} \textbf{\bibinfo{volume}{108}},
  \bibinfo{pages}{041101} (\bibinfo{year}{2012}).

\bibitem[{\citenamefont{Kugo and Ohta}(2014)}]{Kugo:2014aa}
\bibinfo{author}{\bibfnamefont{T.}~\bibnamefont{Kugo}} \bibnamefont{and}
  \bibinfo{author}{\bibfnamefont{N.}~\bibnamefont{Ohta}}
  (\bibinfo{year}{2014}), \eprint{arXiv:1401.3873}.

\bibitem[{\citenamefont{Mirbabayi}(2011)}]{Mirbabayi:2011aa}
\bibinfo{author}{\bibfnamefont{M.}~\bibnamefont{Mirbabayi}}
  (\bibinfo{year}{2011}), \eprint{arXiv:1112.1435}.

\bibitem[{\citenamefont{Berezhiani
  et~al.}(2013{\natexlab{a}})\citenamefont{Berezhiani, Chkareuli, and
  Gabadadze}}]{Berezhiani:2013aa}
\bibinfo{author}{\bibfnamefont{L.}~\bibnamefont{Berezhiani}},
  \bibinfo{author}{\bibfnamefont{G.}~\bibnamefont{Chkareuli}},
  \bibnamefont{and}
  \bibinfo{author}{\bibfnamefont{G.}~\bibnamefont{Gabadadze}},
  \bibinfo{journal}{Phys. Rev. D} \textbf{\bibinfo{volume}{88}},
  \bibinfo{pages}{124020} (\bibinfo{year}{2013}{\natexlab{a}}).

\bibitem[{\citenamefont{Berezhiani
  et~al.}(2013{\natexlab{b}})\citenamefont{Berezhiani, Chkareuli, de~Rham,
  Gabadadze, and Tolley}}]{Berezhiani:2013ab}
\bibinfo{author}{\bibfnamefont{L.}~\bibnamefont{Berezhiani}},
  \bibinfo{author}{\bibfnamefont{G.}~\bibnamefont{Chkareuli}},
  \bibinfo{author}{\bibfnamefont{C.}~\bibnamefont{de~Rham}},
  \bibinfo{author}{\bibfnamefont{G.}~\bibnamefont{Gabadadze}},
  \bibnamefont{and} \bibinfo{author}{\bibfnamefont{A.}~\bibnamefont{Tolley}}
  (\bibinfo{year}{2013}{\natexlab{b}}), \eprint{arXiv:1305.0271}.

\bibitem[{\citenamefont{D'Amico et~al.}(2013)\citenamefont{D'Amico, Gabadadze,
  Hui, and Pirtskhalava}}]{DAmico:2012aa}
\bibinfo{author}{\bibfnamefont{G.}~\bibnamefont{D'Amico}},
  \bibinfo{author}{\bibfnamefont{G.}~\bibnamefont{Gabadadze}},
  \bibinfo{author}{\bibfnamefont{L.}~\bibnamefont{Hui}}, \bibnamefont{and}
  \bibinfo{author}{\bibfnamefont{D.}~\bibnamefont{Pirtskhalava}},
  \bibinfo{journal}{Phys. Rev.} \textbf{\bibinfo{volume}{D87}},
  \bibinfo{pages}{064037} (\bibinfo{year}{2013}).

\bibitem[{\citenamefont{Felice and Mukohyama}(2013)}]{Felice:2013aa}
\bibinfo{author}{\bibfnamefont{A.~D.} \bibnamefont{Felice}} \bibnamefont{and}
  \bibinfo{author}{\bibfnamefont{S.}~\bibnamefont{Mukohyama}},
  \bibinfo{journal}{Physics Letters B 728C (2014), pp. 622-625}
  (\bibinfo{year}{2013}).

\bibitem[{\citenamefont{Gabadadze
  et~al.}(2014{\natexlab{a}})\citenamefont{Gabadadze, Kimura, and
  Pirtskhalava}}]{Gabadadze:2014ab}
\bibinfo{author}{\bibfnamefont{G.}~\bibnamefont{Gabadadze}},
  \bibinfo{author}{\bibfnamefont{R.}~\bibnamefont{Kimura}}, \bibnamefont{and}
  \bibinfo{author}{\bibfnamefont{D.}~\bibnamefont{Pirtskhalava}}
  (\bibinfo{year}{2014}{\natexlab{a}}), \eprint{arXiv:1412.8751}.

\bibitem[{\citenamefont{Horndeski}(1974)}]{Horndeski:1974aa}
\bibinfo{author}{\bibfnamefont{G.}~\bibnamefont{Horndeski}},
  \bibinfo{journal}{Int. J. Theor. Phys.} \textbf{\bibinfo{volume}{10}}
  (\bibinfo{year}{1974}).

\bibitem[{\citenamefont{Deffayet et~al.}(2009)\citenamefont{Deffayet,
  Esposito-Farese, and Vikman}}]{Deffayet:2009aa}
\bibinfo{author}{\bibfnamefont{C.}~\bibnamefont{Deffayet}},
  \bibinfo{author}{\bibfnamefont{G.}~\bibnamefont{Esposito-Farese}},
  \bibnamefont{and} \bibinfo{author}{\bibfnamefont{A.}~\bibnamefont{Vikman}},
  \bibinfo{journal}{Phys. Rev.} \textbf{\bibinfo{volume}{D79}},
  \bibinfo{pages}{084003} (\bibinfo{year}{2009}).

\bibitem[{\citenamefont{Deffayet and Steer}(2013)}]{Deffayet:2013ab}
\bibinfo{author}{\bibfnamefont{C.}~\bibnamefont{Deffayet}} \bibnamefont{and}
  \bibinfo{author}{\bibfnamefont{D.~A.} \bibnamefont{Steer}}
  (\bibinfo{year}{2013}), \eprint{arXiv:1307.2450}.

\bibitem[{\citenamefont{Koyama et~al.}(2013)\citenamefont{Koyama, Niz, and
  Tasinato}}]{Koyama:2013aa}
\bibinfo{author}{\bibfnamefont{K.}~\bibnamefont{Koyama}},
  \bibinfo{author}{\bibfnamefont{G.}~\bibnamefont{Niz}}, \bibnamefont{and}
  \bibinfo{author}{\bibfnamefont{G.}~\bibnamefont{Tasinato}},
  \bibinfo{journal}{Phys. Rev.} \textbf{\bibinfo{volume}{D88}},
  \bibinfo{pages}{021502} (\bibinfo{year}{2013}).

\bibitem[{\citenamefont{Hinterbichler et~al.}(2010)\citenamefont{Hinterbichler,
  Trodden, and Wesley}}]{Hinterbichler:2010aa}
\bibinfo{author}{\bibfnamefont{K.}~\bibnamefont{Hinterbichler}},
  \bibinfo{author}{\bibfnamefont{M.}~\bibnamefont{Trodden}}, \bibnamefont{and}
  \bibinfo{author}{\bibfnamefont{D.}~\bibnamefont{Wesley}},
  \bibinfo{journal}{Phys. Rev. D} \textbf{\bibinfo{volume}{82}},
  \bibinfo{pages}{124018} (\bibinfo{year}{2010}).

\bibitem[{\citenamefont{Padilla et~al.}(2010)\citenamefont{Padilla, Saffin, and
  Zhou}}]{Padilla:2010aa}
\bibinfo{author}{\bibfnamefont{A.}~\bibnamefont{Padilla}},
  \bibinfo{author}{\bibfnamefont{P.~M.} \bibnamefont{Saffin}},
  \bibnamefont{and} \bibinfo{author}{\bibfnamefont{S.-Y.} \bibnamefont{Zhou}},
  \bibinfo{journal}{JHEP} \textbf{\bibinfo{volume}{12}}, \bibinfo{pages}{031}
  (\bibinfo{year}{2010}).

\bibitem[{\citenamefont{Padilla et~al.}(2011)\citenamefont{Padilla, Saffin, and
  Zhou}}]{Padilla:2010ab}
\bibinfo{author}{\bibfnamefont{A.}~\bibnamefont{Padilla}},
  \bibinfo{author}{\bibfnamefont{P.~M.} \bibnamefont{Saffin}},
  \bibnamefont{and} \bibinfo{author}{\bibfnamefont{S.-Y.} \bibnamefont{Zhou}},
  \bibinfo{journal}{JHEP} \textbf{\bibinfo{volume}{01}}, \bibinfo{pages}{099}
  (\bibinfo{year}{2011}).

\bibitem[{\citenamefont{Gabadadze et~al.}(2013)\citenamefont{Gabadadze,
  Hinterbichler, Pirtskhalava, and Shang}}]{Gabadadze:2013aa}
\bibinfo{author}{\bibfnamefont{G.}~\bibnamefont{Gabadadze}},
  \bibinfo{author}{\bibfnamefont{K.}~\bibnamefont{Hinterbichler}},
  \bibinfo{author}{\bibfnamefont{D.}~\bibnamefont{Pirtskhalava}},
  \bibnamefont{and} \bibinfo{author}{\bibfnamefont{Y.}~\bibnamefont{Shang}},
  \bibinfo{journal}{Phys. Rev. D} \textbf{\bibinfo{volume}{88}},
  \bibinfo{pages}{084003} (\bibinfo{year}{2013}).

\bibitem[{\citenamefont{Ondo and Tolley}(2013)}]{Ondo:2013wka}
\bibinfo{author}{\bibfnamefont{N.~A.} \bibnamefont{Ondo}} \bibnamefont{and}
  \bibinfo{author}{\bibfnamefont{A.~J.} \bibnamefont{Tolley}},
  \bibinfo{journal}{JHEP} \textbf{\bibinfo{volume}{1311}}, \bibinfo{pages}{059}
  (\bibinfo{year}{2013}).

\bibitem[{\citenamefont{Gabadadze
  et~al.}(2014{\natexlab{b}})\citenamefont{Gabadadze, Kimura, and
  Pirtskhalava}}]{Gabadadze:2014aa}
\bibinfo{author}{\bibfnamefont{G.}~\bibnamefont{Gabadadze}},
  \bibinfo{author}{\bibfnamefont{R.}~\bibnamefont{Kimura}}, \bibnamefont{and}
  \bibinfo{author}{\bibfnamefont{D.}~\bibnamefont{Pirtskhalava}},
  \bibinfo{journal}{Phys. Rev.} \textbf{\bibinfo{volume}{D90}},
  \bibinfo{pages}{024029} (\bibinfo{year}{2014}{\natexlab{b}}).

\end{thebibliography}

\end{document}